\documentclass[sigconf, review=false, timestamp=false, natbib]{acmart}

\usepackage{hyperref}
\usepackage{amsmath,amssymb,amsfonts}
\usepackage{graphicx}
\usepackage{textcomp}
\usepackage{paralist}
\usepackage[compact]{titlesec}
\titlespacing{\section}{2pt}{*0}{*0}
\titlespacing{\subsection}{2pt}{*0}{*0}

\acmConference{}{}{}

\setcopyright{none}

\usepackage{color}

\newcommand{\citex}[1]{}

\begin{document}

\title{Wrangling Rogues: A Case Study on Managing Experimental Post-Moore Architectures
}

\author{Will Powell}
\email{will.powell@cc.gatech.edu}
\orcid{0000-0002-1570-5930}
\author{Jason Riedy}
\email{jason.riedy@cc.gatech.edu}
\orcid{0000-0002-4345-4200}
\affiliation{\department{School of Computational Science and Engineering}
  \institution{Georgia Institute of Technology}
  \city{Atlanta}
  \state{Georgia}}

\author{Jeffrey S. Young}
\email{jyoung9@gatech.edu}
\orcid{0000-0001-9841-4057}
\author{Thomas M. Conte}
\email{conte@gatech.edu}
\affiliation{\department{School of Computer Science}
  \institution{Georgia Institute of Technology}
  \city{Atlanta}
  \state{Georgia}}

\begin{abstract}
  The Rogues Gallery is a new experimental testbed that is focused on tackling \emph{rogue} architectures for the post-Moore era of computing. While some of these devices have roots in the embedded and high-performance computing spaces, managing current and emerging technologies provides a challenge for system administration that are not always foreseen in traditional data center environments.


We present an overview of the motivations and design of the initial Rogues Gallery testbed and cover some of the unique challenges that we have seen and foresee with upcoming hardware prototypes for future post-Moore research. Specifically, we cover networking, identity management, scheduling of resources, and tools and sensor access aspects of the Rogues Gallery along with techniques we have developed to manage these new platforms.
We argue that current tools like the Slurm scheduler can support new rogues without major infrastructure changes.
\end{abstract}

\maketitle
\pagestyle{plain}

\section{Challenges for Post-Moore Testbeds}
\label{sec:introduction}

As we look to the end of easy and cost-effective transistor scaling, we enter the \emph{post-Moore} era\cite{Vetter_2017} and reach a turning point in computer system design and usage. Accelerators like GPUs have created a pronounced shift in the high-performance computing and machine learning application spaces, but there is a wide variety of possible architectural choices for the post-Moore era, including memory-centric, neuromorphic, quantum, and reversible computing. These revolutionary research fields combined with alternative materials-based approaches to silicon-based hardware have given us a bewildering array of options and \emph{rogue} devices for the post-Moore era. However, there currently is limited guidance on how to evaluate this novel hardware for tomorrow's application needs. 

Creating one-off testbeds for each new post-Moore technology increases the cost of evaluating these new technologies.
We present an in-progress case study of a cohesive user environment that enables researchers to perform experiments across different novel architectures that may be in different stages of acceptance by the wider computing community.
The \emph{\textbf{Rogues Gallery}} is a new experimental testbed focusing on opening access to rogue architectures that may play a larger role in the Post-Moore era of computing.
While some of these devices have roots in the embedded and high-performance computing spaces, managing current and emerging technologies provides a challenge for system administration that are not always foreseen in traditional data center environments.
Our Rogues Gallery of novel technologies has produced research results and is being used in classes both at Georgia Tech as well as at other institutions.

The key lessons from this testbed (so far) are the following: 
\begin{compactitem}
\item \textbf{Invest in rogues, but realize some technology may be short-lived.} We cannot invest too much time in configuring and integrating a novel architecture when tools and software stacks may be in an unstable state of development.  Rogues may be short-lived; they may not achieve their goals.  Finding their limits is an important aspect of the Rogues Gallery.  Early users of these platforms are technically sophisticated and relatively friendly, so not all components of the supporting infrastructure need be user friendly.
\item \textbf{Physical hardware resources not dedicated to rogues should be kept to a minimum.} As we explain in Section \ref{sec:management-issues}, most functionality should be satisfied by VMs and containers rather than dedicating a  physical server to manage each rogue piece of hardware. 
\item \textbf{Collaboration and commiseration is key.}  Rogues need a community to succeed.  If they cannot establish a community (users, vendors interested in providing updates, manageability from an IT perspective), they will disappear. We promote community with our testbed through a documentation wiki, mailing lists, and other communication channels for users as well as close contact with vendors, where appropriate.
\item \textbf{Licensing and appropriate identity management are tough but necessary challenges.}  Managing access to prototypes and licensed software are tricky for academic institutions, but we must work with collaborators like start-ups to provide a secure environment with appropriate separation of privileges to protect IP as necessary.  Many software-defined network (SDN) environments assume endpoint implementations that may not be possible on first-generation hardware.  Documentation restrictions present a particular challenge while trying to build a community.
\end{compactitem}

\begin{figure*}[th]
  \centering
  \hspace*{\fill}
  \includegraphics[width=0.3\textwidth]{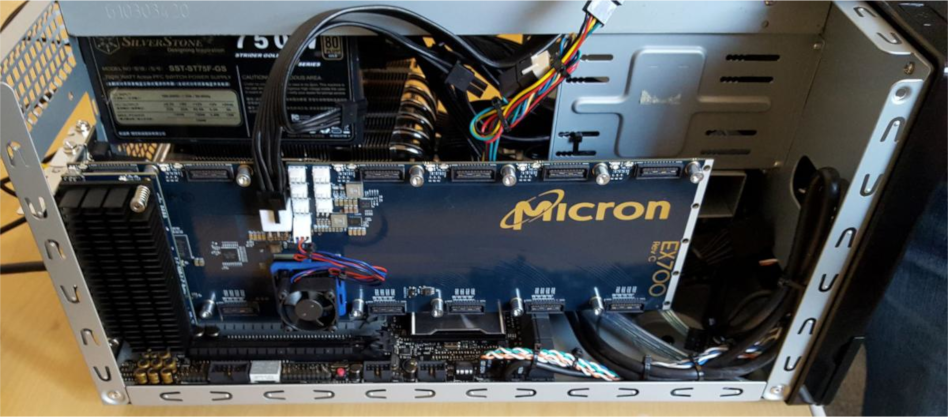}%
  \hspace*{\fill}
  \raisebox{-.3\height}{\includegraphics[width=0.15\textwidth]{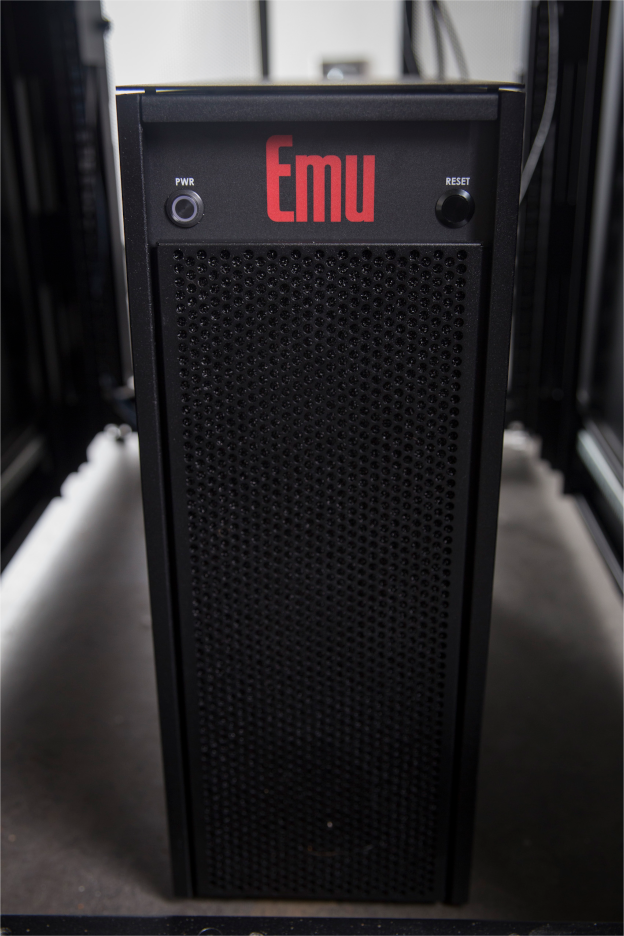}}
  \hspace*{\fill}
  \includegraphics[width=0.3\textwidth]{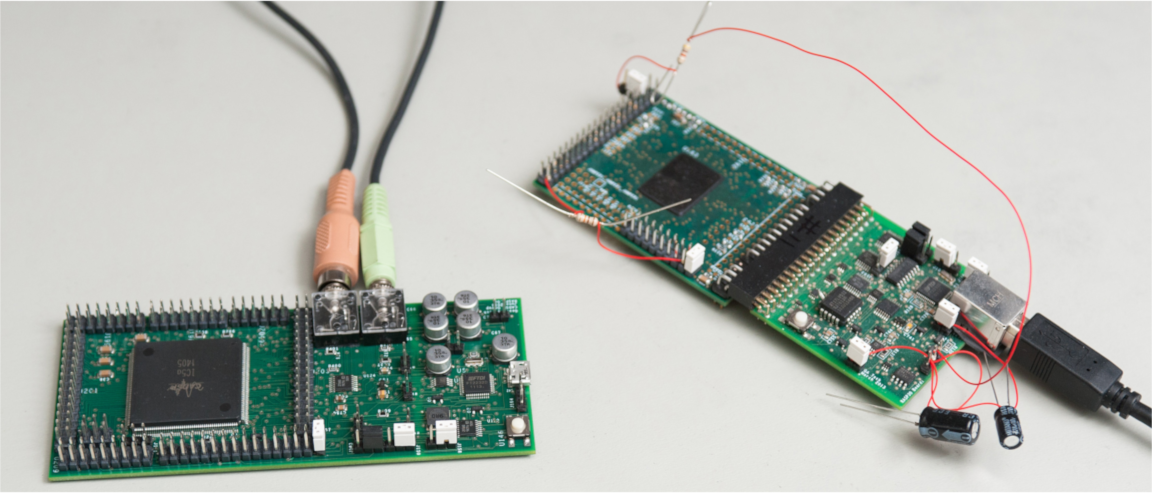}%
  \hspace*{\fill}
  \caption{Initial rogues: FPGAs, the Emu Chick, and the field programmable analog array (FPAA)}
  \label{fig:initial-rogues}
\end{figure*}

Here we describe the initial design of the testbed as well as the high-level system management strategy that we use to maintain and tie together novel architectures.
We use existing tools as much as possible; we need not re-invent the systems integration wheel to support early novel architectures, which may be few in number and specialized.

\section{The Rogues Gallery}\label{sec:rogues-gallery}
The Rogues Gallery was initiated by Georgia Tech's Center for Research into Novel Computing Hierarchies (CRNCH) in late 2017. The Gallery's focus is to acquire new and unique hardware (the rogues) from vendors, research labs, and start-ups and make this hardware widely available to students, faculty, and industry collaborators within a managed data center environment. By exposing students and researchers to this set of unique hardware, we hope to foster cross-cutting discussions about hardware designs that will drive future performance improvements in computing long after the Moore's Law era of cheap transistors ends.  

The primary goal of the Rogues Gallery is to make first-generation start-up or research lab prototypes available to a wide audience of researchers and developers as soon as it is moderately stable for testing, application development, and benchmarking. Examples of such cutting-edge hardware include Emu Technology's Chick, FPGA-based memory-centric platforms, field programmable analog devices (FPAAs), and others.
This mirrors how companies like Intel and IBM are investigating novel hardware, Loihi and TrueNorth respectively, but the Rogues Gallery adds student and collaborator access.

The Rogues Gallery provides a coherent, managed environment for exposing new hardware and supporting software and tools to students, researchers, and collaborators. This environment allows users to perform critical architecture, systems, and HPC research, and enables them to train with new technologies so they can be more competitive in today's rapidly changing job market. As part of this goal, CRNCH and Georgia Tech are committed to partnering with vendors to define a flexible access management model that allows for a functional and distinctive research experience for both Georgia Tech and external users, while protecting sensitive intellectual property and technologies.

Not all rogues become long-term products.  Some fade away within a few years (or are acquired by companies that fail to productize the technology).  The overall infrastructure of a testbed focused on rogues must minimize up-front investment to limit the cost of ``just trying out'' new technology.  As these early-access and prototype platforms change, the infrastructure must also adapt.

And even if the technology is fantastic, rogues that do not develop communities do not last.  In our opinion, initial communities grow by easing use and access. Some systems, like the Emu Chick, require substantial thought around restructuring algorithms and data structures.  Easing access permits eager users (e.g., students) to play with ideas. Georgia Tech has a history of providing such early access through efforts such as the Sony-Toshiba-IBM Center of Competence for the Cell Broadband Engine Processor\cite{sticellcenter},
the NSF Keeneland GPU system\cite{vetter2011keeneland},
and work with early high core count Intel-based platforms\cite{icassp2012-stinger}.

\begin{figure}
  \centering
  \includegraphics[width=0.9\linewidth]{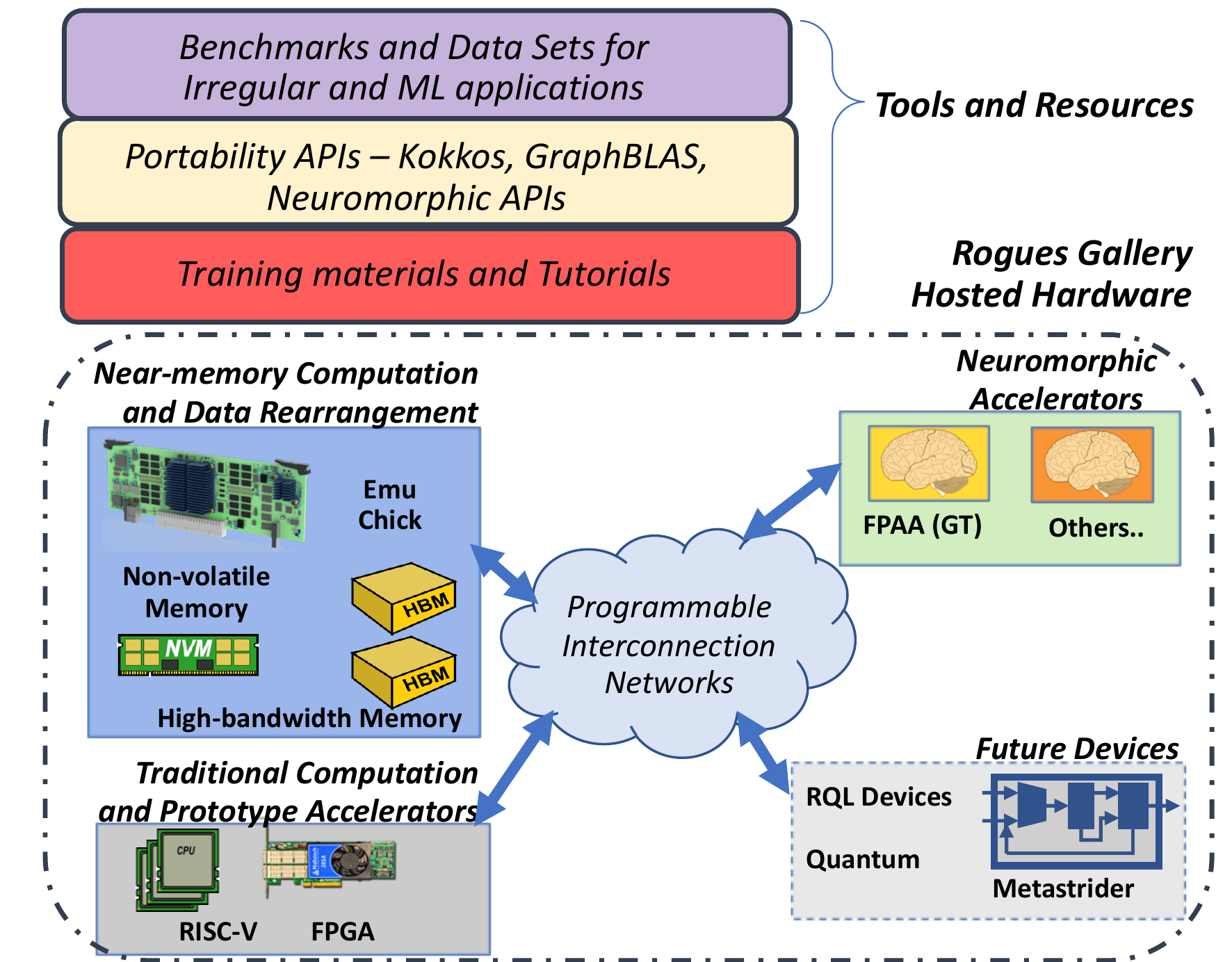}
  \caption{High-level overview of the CRNCH Rogues Gallery}
  \label{fig:rg_infra}
\end{figure}

\section{Initial Rogues}
\label{sec:initial-platforms}

Our initial Rogues Gallery shown in Figure~\ref{fig:initial-rogues} consists of various field programmable gate arrays (FPGAs), an Emu Chick, and a field-programmable analog array (FPAA).  The gallery also currently includes power monitoring infrastructure for embedded-style systems (NVIDIA Tegra), but we will not go into depth on that more typical infrastructure.

\paragraph{FPGAs for Memory System Exploration}

\label{sec:fpgas}

Reconfigurable devices like FPGAs are not novel in themselves, but the gallery includes FPGAs for prototyping near-memory and in-network computing.
Currently the Rogues Gallery includes
\begin{compactitem}
  \item two Nallatech 385 PCIe cards with mid-grade Intel Arria 10 FPGAs and 8 GiB of memory;
  \item a Nallatech 520N PCIe card with a Intel Stratix 10, 32 GiB of memory, and multiple 100 GBps network ports; and
  \item an end-of-lifed Micron AC-510 card that pairs an Xilinx Ultrascale 060 with a 4 GiB Hybrid Memory Cube (HMC).
\end{compactitem}
While now reaching the end of official support, the HMC-based system has many local users and provided results for
several recent publications \cite{hadidi:2017:demysthmc, hadidi:2018:perf-noc-hmc} focused on characterization and utilization of the 3D stacked memory component.
All platforms are used for exploring new FPGA programming paradigms like OpenCL and domain-specific languages.
Related projects like SuperStrider and MetaStrider also are exploring accelerating sparse and graph computations by pushing compute primitives to memory\cite{8123669}.

\paragraph{Emu Chick}

\label{sec:emu-chick}

\begin{figure}
\centering
\includegraphics[width=0.8\linewidth]{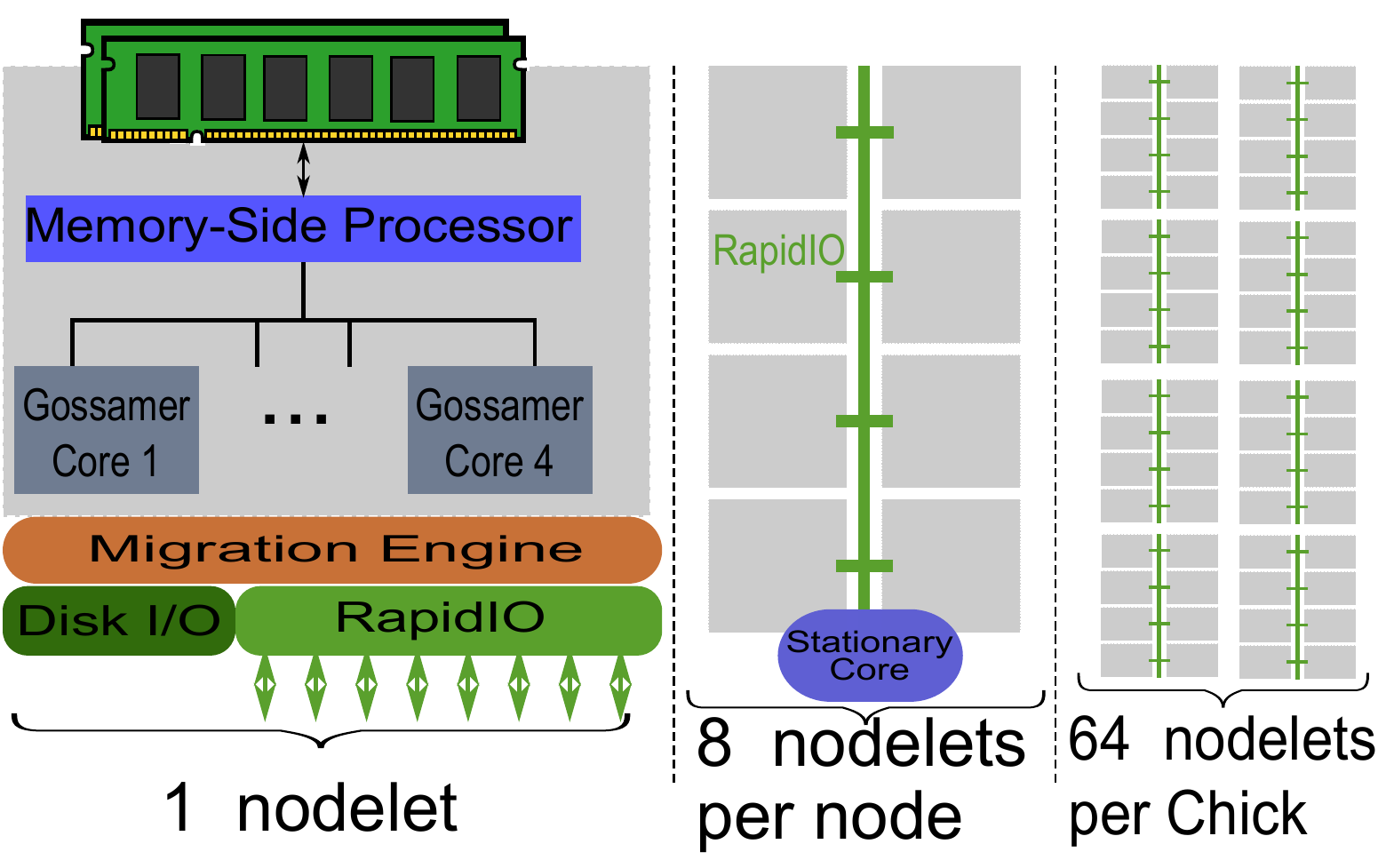}
\caption{{\small{Emu architecture: The system consists of \emph{stationary} processors for running the operating system and up to four \emph{Gossamer} processors per nodelet tightly coupled to memory.}}}\label{fig:emu-arch}
\end{figure}

The Emu architecture (our Emu Chick box is shown in Figure \ref{fig:initial-rogues}) focuses on improved random-access bandwidth scalability by migrating lightweight \emph{Gossamer} threads to data and emphasizing fine-grained memory access.
A general Emu system consists of the following processing elements, as illustrated in Figure~\ref{fig:emu-arch}:
\begin{itemize}
\item A common \emph{stationary} processor runs the operating system (Linux) and manages storage and network devices.
\item \emph{Nodelets} combine narrowly banked memory with several highly multi-threaded, cache-less \emph{Gossamer} cores to provide a memory-centric environment for migrating threads.
\end{itemize}
These elements are combined into nodes that are connected by a RapidIO fabric. The current generation of Emu systems include one stationary processor for each of the eight nodelets contained within a node.
A more detailed description of the Emu architecture is available elsewhere~\cite{dysart2016emu,ashes2018}.
The Emu Chick provides an interesting use case as an intermediate level rogue in that it has a basic Linux-based OS on the stationary cores, but that OS cannot be modified (at time of writing).
Also, development users need administrative access to reset or reconfigure portions of the Chick.
Integrating the Chick into a sufficiently secure infrastructure requires some creativity.

\paragraph{FPAA: Field Programmable Analog Array}
\label{sec:fpaa}
The Field Programmable Analog Array (FPAA)\cite{7374749}, developed at Georgia Tech by Dr.\ Jennifer Hasler's group, is a combined analog and digital board that can implement many analog and neuromorphic architectures\cite{10.3389/fnins.2013.00118,shah:2016:fpaa_acoustic_classifier}.
The FPAA combines a driving 16-bit MSP430 microprocessor with a 2D array of ultra-low power processors consisting of floating-gate analog plus digital blocks.
The exploration and development platform is a USB-attached board with multiple analog input ports (Figure~\ref{fig:initial-rogues}).
All of the development tools for the FPAA and MSP430 are free software based on Scilab and XCos and are distributed in an Ubuntu-based virtual machine image.
The FPAA's 2D array combines floating-gate analog units with programmable digital units along with its routing structure.
Even manufactured with 350nm CMOS, the FPAA uses 23 $\mu$W to recognize the word ``dark'' in the TIMIT database\cite{10.3389/fnins.2013.00118}.
Similarly, classifying acoustic signals from a knee joint requires 15.29 $\mu$W\cite{shah:2016:fpaa_acoustic_classifier}.
Both of these types of computations are performed in real time, so these power savings translate directly to energy savings and justify further research in mixed analog and digital hardware for multiple applications. As the development platform for FPAA is a remote USB device, it provides some challenges in terms of how to monitor, schedule, and maintain it in the Rogues Gallery.

\section{Management Issues}
\label{sec:management-issues}

\begin{figure}[thb]
  \centering
  \includegraphics[width=0.85\linewidth]{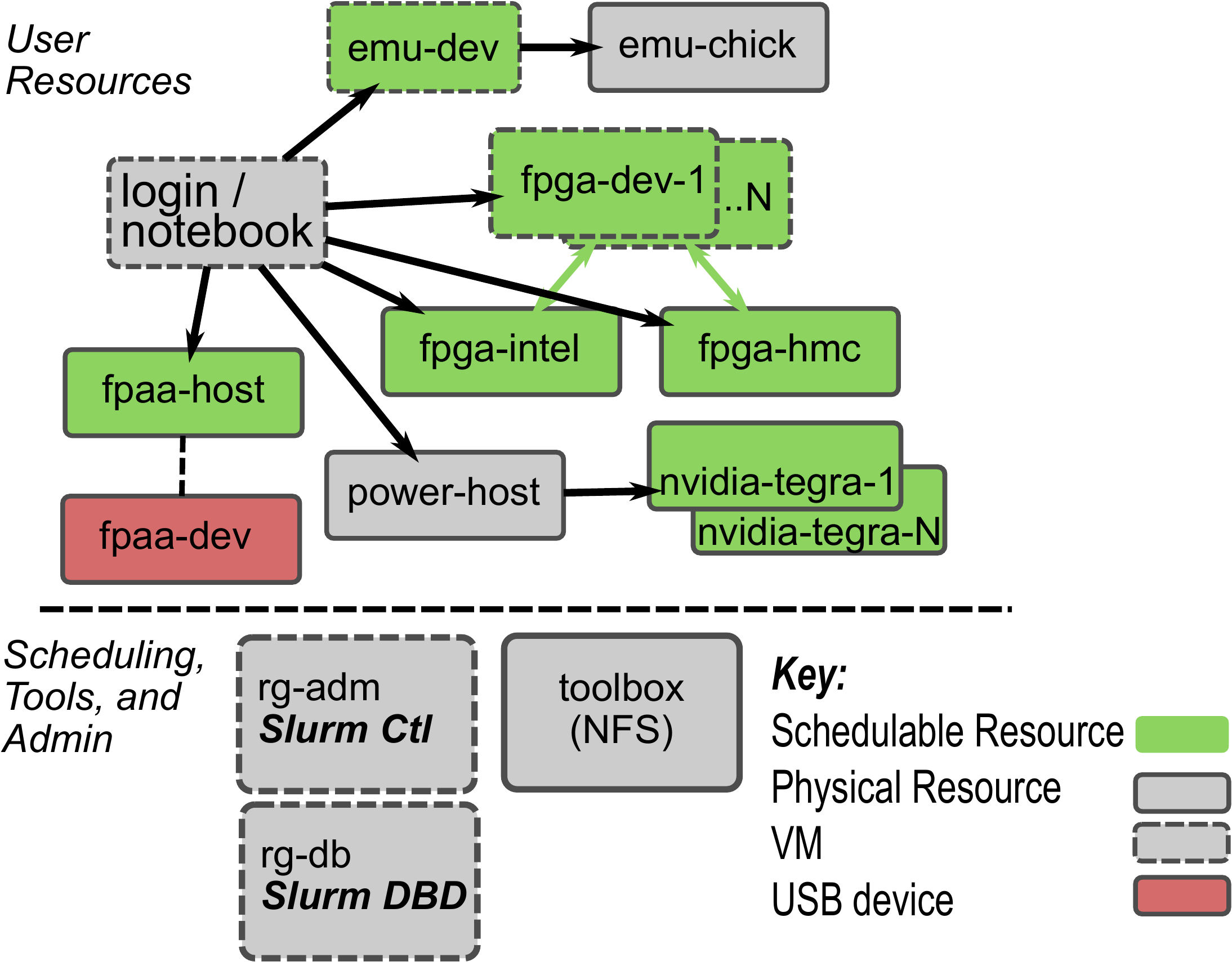}
  \caption{Overview of Rogues Gallery resources and management / network structure}
  \label{fig:rg-overview}
\end{figure}

Figure \ref{fig:rg-overview} shows the management and networking outline of the Rogues gallery. Here we discuss different components of the testbed from the identity management, scheduling, tools support, and networking viewpoints. As demonstrated in the figure, many of the tools and development platforms are hosted on virtual machines while tools are made available via containers where available, as is the case with the Emu Chick simulator and toolchain as well as the FPAA development VM. 

Management of hardware may be complicated because many of the rogues might be temporary additions to the testbed, especially if they fail to build a local userbase or if a company decides to discontinue tool and compiler support. Currently, we plan to reevaluate the hardware composition of the testbed each year and confer with users and industry and national lab colleagues to ensure sysadmin resources are focused on an keeping an up-to-date set of rogues and associated software stacks.

We emphasize using existing tools although sometimes through an extra level of indirection.  For example, we cannot fully integrate the Emu Chick into the resource manager, Slurm\footnote{\url{https://slurm.schedmd.com/}}, at the time of writing since each Chick node's OS image is immutable, but we can manage queue-based access to a VM used to access the Chick.

\subsection{Software Support}
\label{sec:software-development}

Early hardware often is bound to a specific operating system and release.  Tight dependencies make developing and supporting software difficult.  For example, we need compatible versions of Slurm for managing resources, but a specific OS version may not have a compatible version available as a vendor package.  Additionally, some platforms are best used for running code rather than compiling it, like the Chick's stationary cores.  Maintaining many cross-compilation systems for many possible users would defeat our goal of lowering up-front work.

While we may have to support at least two versions of Slurm packages (Redhat and Ubuntu in our case), we find that Singularity, 
a lightweight user-level container system, provides a convenient way to wrap and deploy compilation environments.  For example, we can maintain a single version of tools like the Emu compilers across multiple Ubuntu, Red Hat, \textit{etc}. flavors without overloading the different VMs by sharing minimal development environment containers from a shared tools fileshare across the cluster.  The compilation occurs inside the container as if in the target environment.  With simple wrappers we also can permit use of the images without distributing them if necessary.

We provide Jupyter 
notebooks for external demonstrations and tutorials.  We are working towards leveraging Kata containers\footnote{\url{https://katacontainers.io/}} for limited \emph{untrusted} / anonymous demonstration users.  Data orchestration into and out of the rogues is our primary barrier.

\subsection{Networking Structure}
\label{sec:networking-structure}

The networking structure is designed to protect sensitive intellectual property and licenses as well as provide general security.  Networked devices are attached to firewalling switches that control host and port accesses.  Generally inbound access is allowed in the directions of the arrows in Figure~\ref{fig:rg-overview} and only for ssh.  Outbound access generally is not restricted to allow fetching public data sets from online sources.  The IT administrative systems (not shown) can access everything directly.

One convenient aspect of completely unusual architectures under heavy development, like the Emu Chick, is that they do not present serious attack targets.  Even if someone illicitly accessed the Chick, they would not find many tools to do damage.  However, access is managed to prevent leaking system specifics to inappropriate parties.

\subsection{Identity Management}\label{sec:identity-management}

The Rogues Gallery has two types of identity management that are tied in with its notion of scheduling unique devices. At a high-level, new users request access to the testbed via a publicly available webform tied to the local administrative mailing list. Once approved, users are set up with a new username and given access to specific subsets of resources.
For example, we currently have groups for neuromorphic computing, reconfigurable devices, and the Emu Chick.
Each has sub-groups for normal users and admins (with sudo access to related physical devices and VMs).
This setup integrates with Georgia Tech's campus authentication system (CAS) as well as our robust in-house role (GRS) and entitlement system (GTED).
We leverage Georgia Tech's campus identity system.
Currently the benefits of requiring users to have Georgia Tech logins outweighs the effort in setting up those identities, primarily for access control by POSIX groups and file system permissions.

More challenging is identity management for systems that have a limited idea of what a user might be.
For example, while the Emu Chick currently has a basic Yocto-generated Linux OS\cite{Salvador:2014:ELD:2683875} that runs on the controller node and compute nodes, it does not currently support the subset of packages needed to integrate with our LDAP-based authentication system for users.
When the Emu Chick initially arrived, it only supported root access.  At writing, the Chick still cannot be integrated into an LDAP environment.  This is not uncommon for early hardware.
Likewise, embedded style devices like standalone FPGA boards (such as the Arria 10 devkit) typically run a micro-kernel and have a limited concept of simultaneous user access or limited access to the FPGA fabric.
These devices need to be accessible only through a fully featured front-end that can manage user identity.
In the case of different FPGAs in the same system, this needs coupled to OS-level isolation like Linux control groups.
For machines like the Chick, network firewall rules permit access only from a front-end VM.

\subsection{Scheduling and Management of Resources}
\label{sec:scheduling}

Currently, access to several of the Rogues Gallery resources are maintained using soft scheduling via the group's mailing list and other communication channels.
Some similar external sites use calendaring systems.  Resource slots are represented as conference rooms to be scheduled.
These work well for small numbers of users and resources and require essentially no setup.  The looser mechanisms do require a small, friendly user community.

As our user base grows, we are bringing the resources under control of the Slurm resource manager. 
Slurm already supports generalized resources (GRES) to handle licenses and device allocation.
We use a single Slurm controller for all the devices and rely on features and Slurm's support for heterogeneous jobs\footnote{\url{https://slurm.schedmd.com/heterogeneous_jobs.html}}.
Currently with a small number of each type of hardware, we have not needed to worry about careful queue controls or complex solutions for co-scheduling heterogeneous resources.

Some systems are relatively easy to bring under Slurm control.
The FPGA development and hardware systems are a somewhat typical case of allocating licenses and accelerator nodes.
Our only complication is separating the development machines from the hardware machines; users may need one category or both.
Many researchers start up needing only the compilation and development tools.
Others have their generated FPGA bitstreams and only need the hardware for experiments.
And some will need both for rapid final development turn-around.
The resources need to be scheduled both separately and together, again a use already supported by Slurm.
The \texttt{pam\_slurm\_adopt} module can clean up cross-connections for mutual allocations.
This helps optimize usage for FPGAs and the FPGA development tools.

Other systems like the Emu Chick are more complicated.
While Slurm cannot run on the Emu Chick directly, it can manage allocating the Chick's nodes via generic resources (GRES) on the Emu development node.
Slurm controls access to the front-end VM.
This still requires cooperation between users; we cannot control access to individual nodes without more effort than it is worth.
Also, users need root access to the control node to reboot individual nodes.
Ideally, Slurm could reconfigure the Chick between the eight single nodes and one eight-node configurations, but that is not stable enough to automate.
But we can manage multi-tenant access to the Chick, one user per node, along side single-user access to the entire Chick through different Slurm partitions.  This is similar to sharing a multi-core server while still permitting whole-node jobs.
These are issues with many first-generation hardware platforms.  The risks are worth the benefits in an academic environment.

And then there are USB-connected devices like the FPAAs (Section~\ref{sec:fpaa}), which pose challenges for sharing and scheduling.
Our \emph{plan} is to manage access to the FPAA host (a Raspberry Pi) as with other accelerated systems.
The current tools assume a direct USB connection, and
we are experimenting with tunneling USB/IP\footnote{\url{http://usbip.sourceforge.net/}} from the Pi to the tools running in a VM on the user's machine.
Conveniently, the Pi also supports remote USB power toggling via \texttt{uhubctl}\footnote{\url{https://github.com/mvp/uhubctl}} for hard FPAA resets and provides analog FPAA inputs via a USB digital to analog converter (DAC) of moderate resolution.

\subsection{Tool support}
VMs for tools and users are currently provisioned via a RHEV 4.2 cluster of 16 host nodes, with 4 hosts specifically serving research VMs like those used by the Rogues Gallery. The standard image for most of our testbed research is Ubuntu 18.04.5 LTS with one or two CentOS VMs as needed for specific toolsets.

Early hardware may require its own, specific OS and dependency stack.  New hardware companies rarely can invest in wide software support.
Virtual machines and light-weight containers come to the rescue.
We use Singularity\cite{Kurtzer_2017} to package the Emu toolset and VirtualBox\cite{oai:cds.cern.ch:1641847} for the FPAA tools.

Singularity wraps the Emu compiler and simulator in a manner convenient for command-line build systems.
We can offload compilation and simulation to laptops and other platforms that are less resource-constrained than the VMs.
Only users with access to the Emu development node have access to the Singularity image.
Because of Emu's rapid development, we do not worry about old versions in the wild after students graduate; they provide no useful information outside of what is publicly available.

The Georgia Tech FPAA tools\cite{jlpea6010003} currently consist of a graphical environment using Scilab and Xcos.
These need USB access to the FPAA, so an environment like VirtualBox is more appropriate to encompass the GUI, compiler tools, simulation, and example code.  As mentioned in Section~\ref{sec:scheduling}, we are experimenting with methods for remote USB access.
Here a major unsolved aspect is providing appropriate physical test inputs to an FPAA device as has been done in previous work focused on acoustic classifiers \cite{shah:2016:fpaa_acoustic_classifier}. While we can replay existing audio samples into the FPAA via a DAC converter, we would like to eventually enable an experimental station where students can work with real-world inputs such as environmental sounds.

We manage FPGA and other licensed tools on a mid-range, dual-socket Intel Xeon system (IBM System x3650 M4) with 192GiB of RAM, a 250GB NVMe drive, and a 500GB ZFS raidz-1 volume.
Resources remotely NFS mount the specific tools using autofs.
This does require administrative support for deploying new tools, but managing a single shared host is simpler than managing tools on each individual resource. Currently tool versioning is provided by a top-level symbolic link in each resource's top-level directory while keeping older versions accessible when useful (e.g., \textit{/usr/local/emu} points to the current Emu compiler toolchain).

\subsection{Monitoring the Rogues Gallery}\label{sec:monit-rogu-gall}

An instance of OpenNMS monitors the Rogues Gallery. Alerts are sent to support personnel regarding any change in the availability of systems in case something unexpected/unscheduled occurs. We also use monitored power distribution units for reports/trends as well as alerts if power usage gets dangerously high (intentionally or not).

However, things are complicated with some of the more novel systems in the Rogues Gallery. For example, with the Emu Chick, we would like to monitor the system management board but not be notified every time a user resets an internal node unless there is an issue. We are investigating how to use tools like ssh-ping to supplement OpenNMS queries for the management node without interfering with other system uses.
The FPAA provides a similar challenge in that it is a USB device attached to a related physical host. We would like  to check the USB device's online status without interfering with any active activity and also communicate the availability of the resource to our Slurm scheduler. Our current target for performing this type of monitoring is to extend LBNL's Node Health Check \cite{nhc:gitrepo} script for Slurm to support monitoring (and possibly restarting) specific USB devices.

Monitoring individual FPGA resources without disturbing running experiments similarly is complicated. Currently these devices must be managed as either USB or PCIe devices using node-health scripts. Platforms with an embedded ARM core like Xilinx's Zync board or the Intel Arria10 DevKit provide a basic Linux-enabled core, but it is not clear that these on-board processors can host meaningful monitoring and/or perform communication with a global monitor or scheduler. 

\section{Reproducibility and Replicability}
\label{sec:repr-repl}

Many major research venues push for reproducible, or at least replicable, experiments.
In general this is a wonderful direction.
For early, novel platforms, this may not be easily achievable.
Some of our platforms, like the Emu Chick, fix show-stopping bugs with frequent software and firmware releases. Reverting to earlier versions to reproduce experiments requires system-wide pain and in many cases is just not feasible.

Other Emu Chick installations have used GT-developed benchmarks and test codes like Emu-STREAM and pointer chasing benchmarks\footnote{\url{https://github.com/ehein6/emu-microbench}}, and this has helped Emu Technology identify hardware and software issues.  Other test codes that have been developed for the Rogues Gallery are not generally available do to their one-off nature for a specific hardware, firmware, and experimental setup that has changed drastically during the hardware's deployment. With the Emu, the platform's software interface and usage API is changing as well, in part due to research undertaken over the past year using the Rogues Gallery and due to vendor upgrades.

The high-initial-effort ``Review of Computational Results Artifacts'' option in~\cite{Heroux:2015:EAT:2786970.2743015} still is possible and worthwhile.
Instilling these ideas in students will pay off over the long run but does require more initial effort than is generally allowed for with academic experimentation and publication cycles. This balancing act for replicating results and meeting deadlines with constantly evolving hardware is an open issue for further research.

\section{Education, Outreach, and Collaboration}
\label{sec:education-outreach}

One benefit to hosting the Rogues Gallery at a university is integrating the Gallery into education.
Our initial undergraduate research class, part of the Vertically Integrated Projects program\cite{asee_peer_28697}, provides novel architecture access for early computing and engineering students\footnote{\url{http://www.vip.gatech.edu/teams/rogues-gallery}}.
The students are engaged and self-organized into groups focused on the FPAA, the Emu Chick, and integrating quantum simulators like Qiskit\cite{Qiskit}.
These are students with little initial knowledge of parallel computing, but we hope the experience with novel architectures will prepare them for a wider, post-Moore computing landscape.

As stated previously, no rogue can survive without a community.
People need to learn about the platforms and kick the platforms' tires.
One mechanism is through organized tutorials.
In April 2018 we held a neuromorphic workshop combining invited speakers with hands-on FPAA demonstrations.
In April 2019, we presented a tutorial at ASPLOS\cite{asplos19-tutorial} focused on optimizing for the Emu Chick.
And in July 2019, we presented a similar tutorial at PEARC\cite{pearc19-tutorial}.
We also organize sessions at scientific computing conferences that bring together potential users, the Rogues Gallery, and other test beds.
Tutorial and presentation materials are made available through our external site \url{https://crnch-rg.gitlab.io} and official center site \url{https://crnch.gatech.edu}.

Another way to build community is through collaboration with existing high-performance computing and data analysis groups.
There are active projects to explore both Kokkos\cite{CarterEdwards20143202} and the Graph\-BLAS\cite{kepner16_mathem_found_graph} on the Emu Chick.
The Emu PGAS model is sufficiently unusual that these are not direct ports but re-working of implementation layers to match the architecture.
The FPAA and other neuromorphic-capable platforms present programming challenges that are being addressed by collaborators, such as researchers at UT Knoxville working on the TENNLab framework\cite{psb:18:ten}. We also anticipate a growing collaboration between other Department of Energy (DoE) architecture test beds including CENATE\cite{Tallent_2016} at Pacific Northwest National Lab, ExCL\footnote{\url{https://excl.ornl.gov/}} at Oak Ridge National Lab, and Sandia HAAPS\footnote{\url{https://www.sandia.gov/asc/computational_systems/HAAPS.html}}. With a limited amount of personnel and funding resources to tackle post-Moore computing hardware and research, we believe that each of these centers can help to fulfill different but overlapping research agendas and can coordinate commonalities within academic, industry, and government userbases.

\section{Future Plans}
\label{sec:future-plans}

Clear future plans include extending the infrastructure and making new system integration easier.  Collaborating with other rogue-like centers, including CENATE and ExCL could expose abstractions useful across all such efforts including common scheduling and user management techniques that can also preserve security and data protections. Additionally, solutions to the monitoring issues in Section~\ref{sec:monit-rogu-gall} are crucial for enabling research for a primarily remote userbase. Finally, we must collaborate to establish community standards about reproducibility for changing (and possibly disappearing) novel computing platforms.

We are currently looking to adopt new platforms and combine them with upcoming technology like large-scale nonvolatile memory (\textit{e.g.} Intel Octane DC\cite{izraelevitz19_basic_perfor_measur_intel_optan}) and programmable networking, as shown in Figure \ref{fig:rg_infra}.
A recent NSF award at Georgia Tech for a general-purpose CPU and GPU cluster with XSEDE\cite{Towns_2014} integration opens more pathways for exposing the Rogues Gallery infrastructure to a wider community through common user authentication and advertising the Rogues Gallery as a community resources. Integrating the Globus tools with the rogue architectures will be an interesting challenge but will permit larger scaling and data movement to the host VMs and shared storage that support tools and front-end testing.

\textit{Handling of Sensitive Data Sets} Many interesting applications for these novel platforms involve sensitive data in areas like
corporate networks\cite{Senator:2013:DIT:2487575.2488213} and health care\cite{chai-duke}
A secure infrastructure to pipe sensitive data through our testbed would benefit both application and hardware / software designers.
In this environment, application designers can learn what upcoming hardware could be useful while hardware and software designers would discover which of their assumptions apply to specific, real-world application areas.

Current VLAN, VXLAN, and overlay network methods combined with some assurance for wiping data may suffice for non-regulated industry applications.
Per-VM memory and storage encryption combined with the Science DMZ design\cite{sciencedmz} could assist with data transfer.
However, for some levels of certification, particularly in health care, this may not be feasible.
Identifying specific roadblocks could help new platform developers engineer around them and possibly provide regulatory agencies guidance otherwise unavailable.

\textit{Future Post-Moore Hardware}  We also have not discussed the deployment of \emph{far-off} devices like those in the quantum computing space in our current testbed.
There are upcoming devices in the Noisy Intermediate-Scale Quantum (NISQ)~\cite{Preskill2018quantumcomputingin} category which we are investigating.
Many existing quantum devices require extensive physical hosting requirements that are out of reach for nearly all facilities. However, we envision hosting common sets of quantum tools like those provided by the ProjectQ team \cite{steiger2018projectq} and engaging with smaller quantum start-ups to provide remote access for researchers, perhaps allowing our testbed to become a gateway for this hardware or even a \emph{Rogues Grid}.

\section{Summary}
\label{sec:summary}

The CRNCH Rogues Gallery at Georgia Tech lets researchers experiment with new and novel architectures in a managed but evolving testbed environment.
Reducing the intellectual cost of trying new systems enables new and sometimes unexpected applications that can be mapped onto a post-Moore platform. This low barrier to experimentation is supplemented by a growing community that can help with prototyping and porting software and that can help to give vendors feedback on developing new post-Moore hardware.

Although there are multiple vectors for growth and improved infrastructure with the Rogues Gallery, the testbed has already led to some early successes. These positive outcomes include several published academic papers \cite{young18:microb_charac_emu_chick,hein18_progr_strat_irreg_algor_emu_chick,hadidi:2018:perf-noc-hmc,hadidi:2017:demysthmc,8123669}, support for PhD thesis research, ongoing collaborations with external academic, industry, and government users, and a job offer from one of the rogue startups for at least one of our PhD students at Georgia Tech. We look forward to new infrastructure developments, student-focused activities like the Rogues Gallery VIP class, and further collaborations with other post-Moore architecture and system evaluation labs to help drive the next phase of the Rogues Gallery's evolution.

\begin{acks}
This work is supported in part by Micron's and Intel's hardware donations, the \grantsponsor{NSF}{NSF}{https://nsf.gov} XScala (\grantnum{NSF}{ACI-1339745}) and SuperSTARLU (\grantnum{NSF}{OAC-1710371}) projects, and \grantsponsor{IARPA}{IARPA}{https://iarpa.gov}.
\grantsponsor{SNL}{Sandia National Laboratory}{https://sandia.gov} provides student support for the Rogues Gallery VIP undergraduate research class.
Thanks also to  colleagues at GTRI including David Ediger and Jason Poovey and users like Vipin Sachdeva for ideas and assistance with scheduling and FPGA support using Slurm. In addition, thanks to Eric Hein, Janice McMahon and the rest of the team at Emu for their assistance in setting up and supporting the Emu Chick system for our users.  We thank reviewers for their attentive comments.

\end{acks}
\bibliographystyle{ACM-Reference-Format}
\bibliography{bib/rg.bib}

\end{document}